\documentclass[twoside]{article}

\usepackage{lipsum} 

\usepackage[sc]{mathpazo} 
\usepackage[T1]{fontenc} 
\usepackage{microtype} 
\usepackage[hmarginratio=1:1,top=32mm,columnsep=20pt]{geometry} 
\usepackage{graphicx}
\usepackage{float}
\usepackage{multicol} 
\usepackage[hang, small,labelfont=bf,up,textfont=it,up]{caption} 
\usepackage{booktabs} 
\usepackage{float} 
\usepackage{hyperref} 
\usepackage{amsmath}
\usepackage{lettrine} 
\usepackage{paralist} 
\usepackage{pdfpages}

\usepackage{abstract} 

\usepackage{titlesec} 
\renewcommand\thesection{\Roman{section}} 
\renewcommand\thesubsection{\Roman{subsection}} 
\titleformat{\section}[block]{\large\scshape\centering}{\thesection.}{1em}{} 
\titleformat{\subsection}[block]{\large}{\thesubsection.}{1em}{} 

\usepackage{authblk}

\newenvironment{Figure}
  {\par\medskip\noindent\minipage{\linewidth}}
  {\endminipage\par\medskip}

\usepackage{fancyhdr} 
\title{\vspace{-15mm}\fontsize{24pt}{10pt}\selectfont\textbf{Laser pulse propagation in a meter scale rubidium vapor/plasma cell in AWAKE experiment}} 

\author[1, 3]{A. Joulaei}
\author[1]{J.Moody}
\author[2]{N. Berti}
\author[2]{J. Kasparian}
\author[3]{S. Mirzanejhad}
\author[1]{P. Muggli}
\affil[1]{ \small Max-Planck Institute for Physics, Munich, Germany,}
\affil[2]{University of Geneva, Switzerland,}
\affil[3]{Uiversity of Mazandaran, Iran.}

\vspace{2mm}

\date{November 2015}

\begin{document}

\maketitle 
\thispagestyle{fancy} 


\begin{abstract}
We present the results of numerical studies of laser pulse propagating in a 3.5~cm Rb vapor cell in the linear dispersion regime by using a 1D model and a 2D code that has been modified for our special case. The 2D simulation finally aimed at finding laser beam parameters suitable to make the Rb vapor fully ionized to obtain a uniform, 10~m-long, at least 1~mm in radius plasma in the next step for the AWAKE experiment.

\end{abstract}


\begin{multicols}{2} 

\section{Introduction}

\lettrine[nindent=0em,lines=3]{}
AWAKE is a proof-of-principle proton-driven plasma wakefield electron acceleration experiment at CERN \cite{AwakeReport,PhysicsAwake}. The purpose of the experiment is to understand the physics of the acceleration process to demonstrate high-gradient acceleration 
of electrons from MeV to GeV energy scale within a 10 meter plasma with an electron density of $10^{14}-10^{15}~cm^{-3}$ and driven by a proton bunch.%
The experimental program is based on the numerical simulations of the beam and plasma interaction.%
An ultrashort terawatt Gaussian laser pulse ionizes  a rubidium (Rb) vapor in a 10~m long vapor/plasma cell, creating plasma that acts as an energy transformer from the proton drive bunch to the electron witness bunch that accelerates.%
It also seeds the self-modulation instability with a sharp, moving ionization front.%
 To mitigate the effects of plasma density ramps at the entrance and exit of the plasma, limiting apertures have been proposed. 
The minimum aperture radius is limited by diffraction of the laser pulse when crossing the entrance one. Also, along propagation of the laser pulse in the gas/vapor to ionize and in the plasma, dispersion, kerr-induced self-focusing and filamentation may occur. These effects potentially lead to a decrease in laser pulse intensity below the rubidium ionization threshold, thereby limiting the radial and longitudinal extent of the plasma. %
Some of these effects may be particularly important when the laser pulse spectrum overlaps with an atomic transition of the medium to ionize and the pulse experiences anomalous dispersion effects, as in the AWAKE Rb vapor case. 

For the AWAKE experiment, the 
propagation of the laser pulse intensity is simulated by using a 2D code \cite{gasFilamentation,gasFilamentationthesis}. We initially test the code in the low intensity regime of laser pulse and compare results between 1D calculations, results of the 2D code and also with 
experimental results. To ensure the code accuracy in the %
high-intensity laser regime
, full ionization of the vapor and plasma creation

In this paper, we present results of numerical studies of laser pulse intensity after propagation in 
 a 3.5~cm-long Rb vapor cell in linear regime. 
 


\section{
Simulations}
 The laser pulse interaction with the Rb vapor 
 is simulated with a code that is typically used to model the ionizing laser propagation through the atmosphere \cite{gasFilamentationthesis}. The code is 2D 
 with cylindrical coordinate (r,z). 
It solves the master paraxial propagation equation:  
\begin{equation}
\bigtriangledown^{2}E -  \frac{1}{c^2} \partial_{t}^{2} E = \mu_0 (\partial_t J + \partial_{t}^2 P).
\label{equ:propagation equation}
\end{equation}
\hspace{1mm}
Here E is the laser pulse electric field profile, J is the current and P is the polarization, including the linear %
 and nonlinear response of the medium. 
 The polarization is written in orders of E as:
\begin{equation}
	P(t)=\varepsilon _{0}(\chi ^{1} E(t)+\chi ^{2} E^{2}(t)\ + ...).
	\label{equ:nonlinear polarization}
\end{equation}
\hspace{1mm}
The coefficients $\chi^{n}$ are the n-th order susceptibilities of the medium and the presence of such a term is generally referred to as an n-th order non-linearity.
Typically, 
\autoref{equ:propagation equation} 
is solved also to all orders in the 2D code %
by using the Fourier transform split step method. The equation is solved in the frequency and wave number domain and a unitary transformation (phase shift transformation) is applied with the resulting 
longitudinal wavenumber $k_z$. This transformation is responsible for diffractive and linear dispersion effects.
 In this method 
 the linear dispersion and diffraction effects are applied in the frequency and wave number domains. 
 We use the Hankel-Fourier transform:
 \begin{equation}
	F(k) = \int_{r=0}^{\infty}\int_{\theta =0}^{2\pi}f(r,\theta )e^{ikrcos(\theta )}rdrd\theta.
 \end{equation}
 \hspace{1mm}

when cylindrical symmetry applied to this, it becomes:

 \begin{multline}
  F(k)=\int \int f(r,\theta )e^{ik.r}rdrd\theta\
  \\=2\pi\int_{0}^{\infty}f(r)J_{0}(kr)rdr.
  \label{equ:Hankel-Fourier transform with cylindrical symmetry}
    \end{multline}
Its inverse is: 
\begin{multline}
 f(r)=\frac{1}{(2\pi)^{2}}\int F(k)e^{-ik.r}dk=\
 \\ \frac{1}{2\pi}\int_{0}^{\infty}F(k)J_{0}(kr)rdr.
 \label{equ:inverse Hankel-Fourier transform}
\end{multline}
\hspace{1mm}
The Hankel-Fourier transform of order zero is essentially the 2-dimensional Fourier transform of a circularly symmetric function.
 The code changes 
 between the frequency and time domains. The laser pulse is implemented in the time domain %
 in order to apply intensity dependent effects and the nonlinear polarization terms as well as ionization losses and kerr effect.  
\begin{Figure}
     \centering
     \includegraphics[trim=1cm 16cm 0cm 2cm, clip=true,scale=0.45]{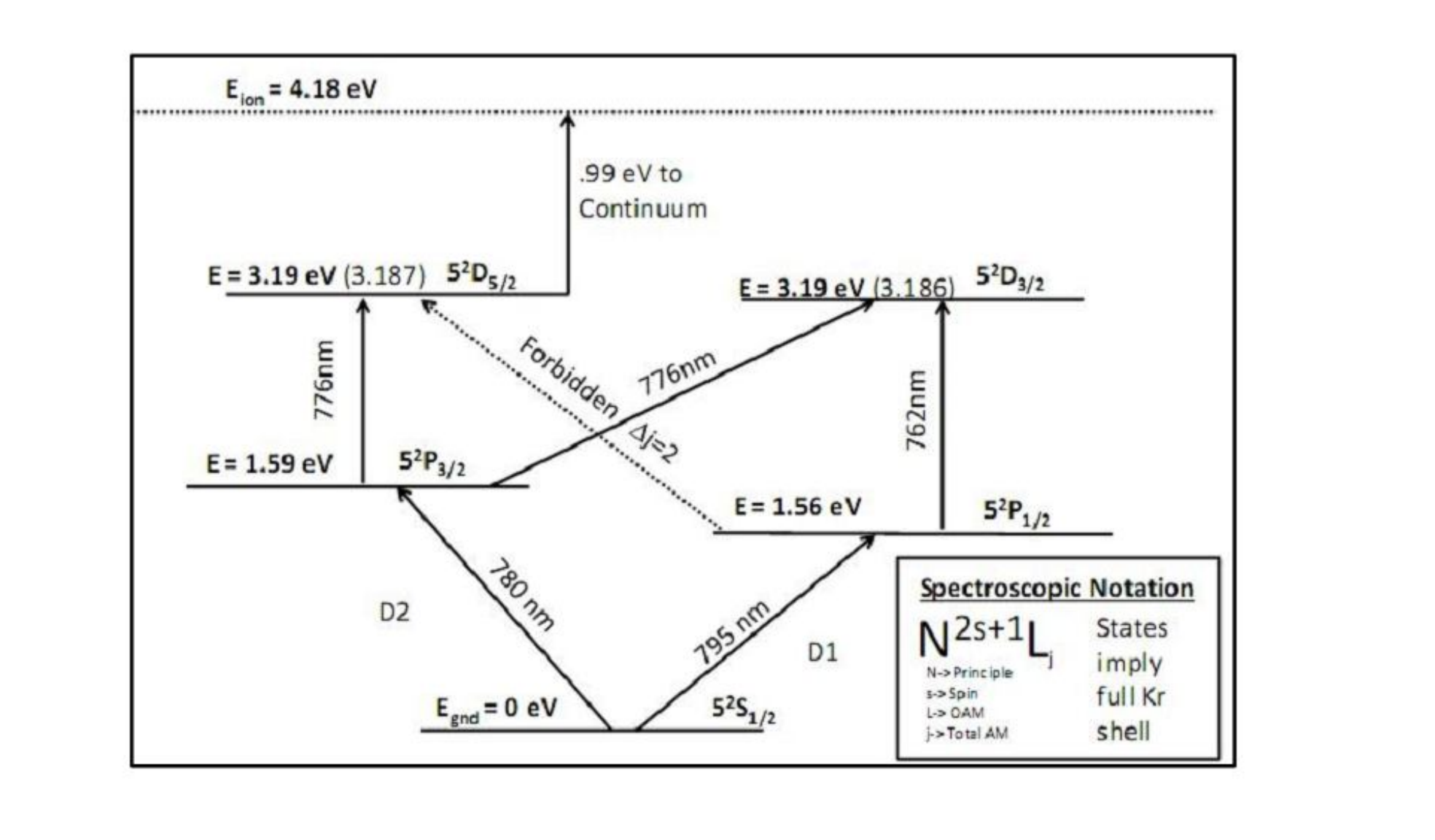}
	 \captionof{figure}{Diagram of Rb valence electronic state \cite{RbTransitionLines}.}
     \label{fig:Diagram of Rb valence electronic state}
 \end{Figure}
\hspace{1mm}  

\begin{Figure}
     \centering
     \includegraphics[trim=3cm 18cm 1cm 1cm, clip=true,scale=0.55]{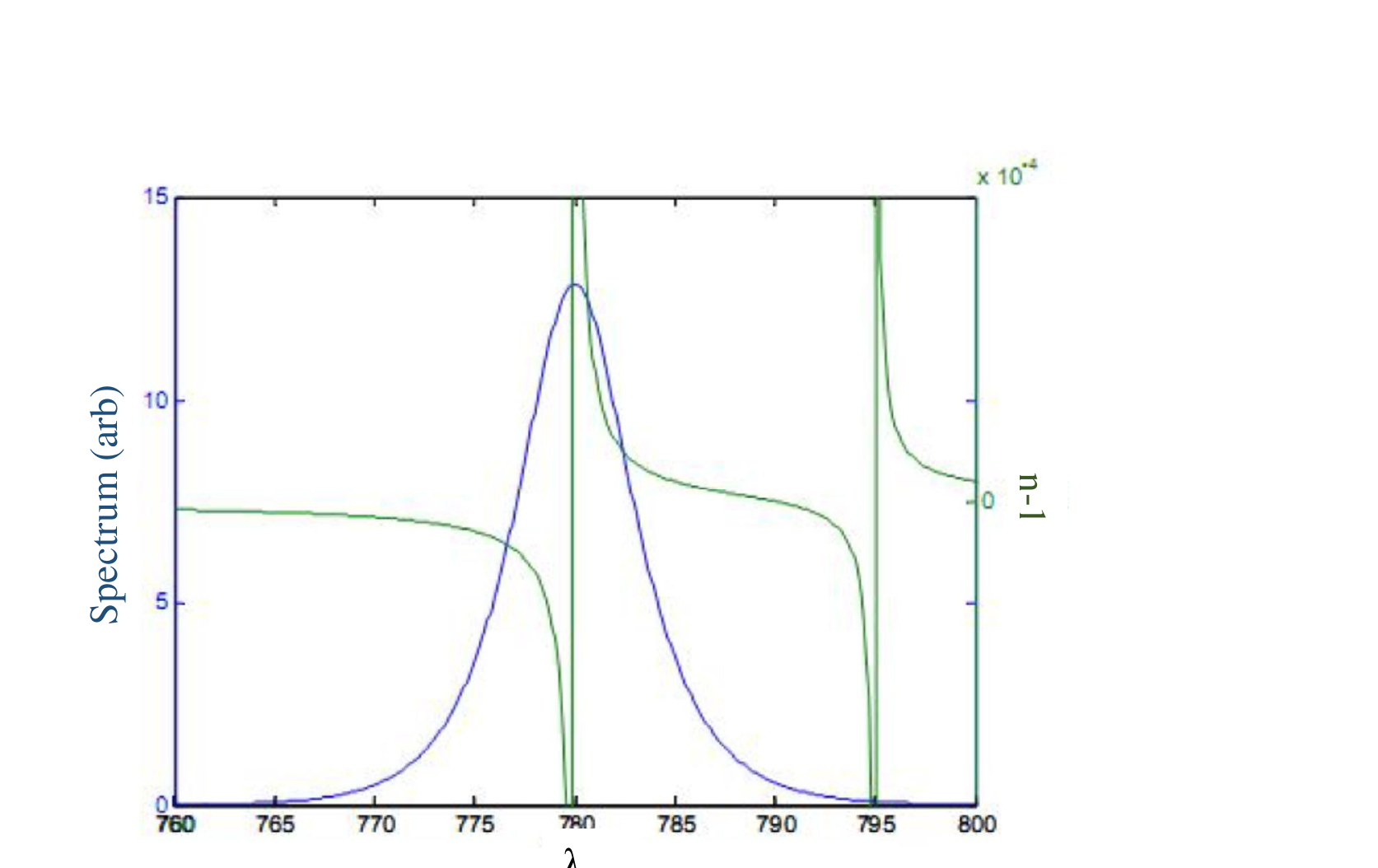}
	 \captionof{figure}{
	 Laser spectrum (blue line) and real part of the refractive index (green line) of Rb near the D2 optical transition line from the ground state with density $6\times10^{14}~cm^{-3}$.}
     \label{fig:spectrumrb}
 \end{Figure}



\begin{Figure}
     \centering
     \includegraphics[trim=5cm 16cm 0.5cm 1cm, clip=true,scale=0.65]{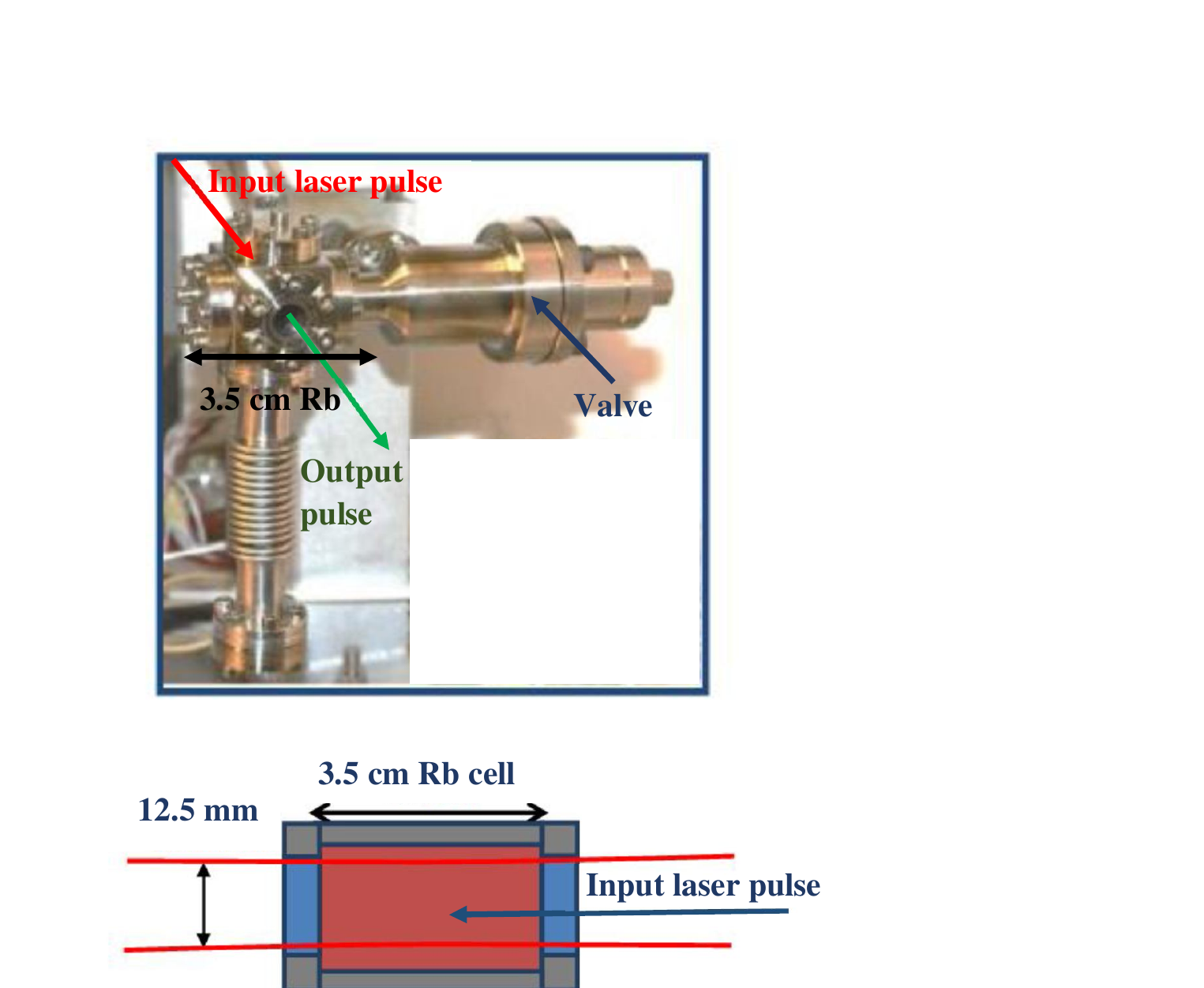}
	 \captionof{figure}{
	 Scheme of the 3.5~cm Rb source}
     \label{fig:3.5cm Rb}
 \end{Figure}
\hspace{5mm} 
The linear interaction between light at frequency $\omega_0$ with the medium with atomic absorption lines at $\omega_{ij}$ 
is described by the linear susceptibility \autoref{equ:linear chi} \cite{NonlinearOptics}:
\begin{equation}
\chi ^{1}(\omega )=\sum _{i}\frac{N_ie^{2}}{\varepsilon _{0}m_{e}}\sum _{j}\frac{f_{ij}}{\omega _{ij}^{2}-\omega _{0}^{2}+i\Gamma _{i}\omega _{ij}}.
\label{equ:linear chi}
\end{equation}
The coefficients $\chi^{1}$ is the linear susceptibility, $N_{i}$ the atomic density, $e$ the electron charge, $\varepsilon_{0}$ vacuum permittivity, $m_{e}$ is the electron mass, $f_{ij}$ absorption oscillator strength, $\omega_{ij}$ is the absorption line of vapor, $\omega_{0}$ central laser frequency and $\Gamma_{i}$ is the transition life time. This parameter describes the linear polarization as the linear 
 response of the medium since:
\begin{equation}
P^{1}(t)=\varepsilon _{0}\chi ^{1}E(t).
\label{equ:linear polarization}
\end{equation}
Based on the two equations below 
 the refractive index and wave vector are also calculated in the frequency domain of the 2D code:
\begin{equation}
n%
_1(\omega )=\sqrt{1+\chi ^{1}(\omega )}.
\label{equ:refractive index}
\end{equation}

\begin{equation}
k(\omega)=n(\omega )\frac{\omega }{c}.
\label{equ:wavenumber}
\end{equation}



\section{Methods}


The laser used in the experiment is %
an Erbium-doped fiber oscillator with a central wavelength of 780 nm 
and a bandwidth of approximately 10 nm. The laser 
is a chirped pulse amplification system that stretches the $\sim100$ fs  pulse to $\sim 100$ ps, amplifies it to a maximum 
energy of 600 mJ, then recompresses 
it to a peak power of $\sim4$ TW at a maximum energy of 450 mJ. 
The plasma is produced from a $%
1-10^{15}~cm^{-3}$ rubidium vapor source. Rubidium was chosen over other 
alkali metals 
because of its low ionization 
potential of 4.2 eV and the relative technical simplicity of its use%
, including producing the baseline vapor density at 200
~$^o$C
. It is also solid at room temperature \cite{RbVaporsource}. 
To create the plasma, the rubidium is photo-ionized by 
multiphoton ionization (AC field ionization) described by %
the Keldysh parameter \cite{KeldyshTheory,Ionization}. Because we are operating in a low Keldysh parameter regime in which the intensity is high and frequency low, the ionization time is much less than the laser pulse length and occurs at a laser threshold intensity of %
$\sim2$ TW/cm$^{2}$.


We use a small rubidium cell with a length of 3.5~cm and the diameter of 12.5 mm 
and a vapor density of $6%
\times10^{14} ~cm^{-3}$. Upon verification of the rubidium vapor density for a given temperature, 400 $\mu J$%
~of laser pulses with the pulse width FWHM 
of 100 fs and an average intensity of 1~$GW/cm^{2}$ 
is propagated through the cell. 
In this low intensity regime the vapor ionization is negligible.

Figure 1
~shows some of the atomic transitions of the Rb atom.
In patricular, Rb has a transition (known as D2) from the ground states at 780.241~nm that falls near the peak of the laser pulse spectrum. \autoref{fig:spectrumrb}  shows the anomalous dispersion in the vicinity of this line calculated using \autoref{equ:linear chi}. This anomalous dispersion can stretch the laser pulse.

\hspace{2mm}


\section{Result and analysis}


The length of the laser pulse is determined 
with a second harmonic generation 
intensity autocorrelator 
. As a background, the pulse length is measured through the cell at room temperature %
(no Rb) and as expected, there is negligible pulse stretching through the cold cell. This can be seen from autocorrelation data of the experiment that is shown on \autoref{fig:Laser pulse intensity auto-correlation simulation}. 
The same figure shows that over the 3.5~cm cell %
with Rb density of $6\times10^{14} ~cm^{-3}$ the pulse stretches dramatically 
with the pedestal increasing in amplitude by a factor of 5. 

 \autoref{fig:Propagated laser pulse intensity simulation} shows the laser pulse intensity simulation 
from the 1D model and 2D code 
after propagation through the 3.5~cm %
Rb cell. The models are consistent with each other and the two dimensional numerical effects are small. 

\begin{Figure}
     \centering
     \includegraphics[trim=3cm 14cm 2cm 2cm, clip=true,scale=0.45]{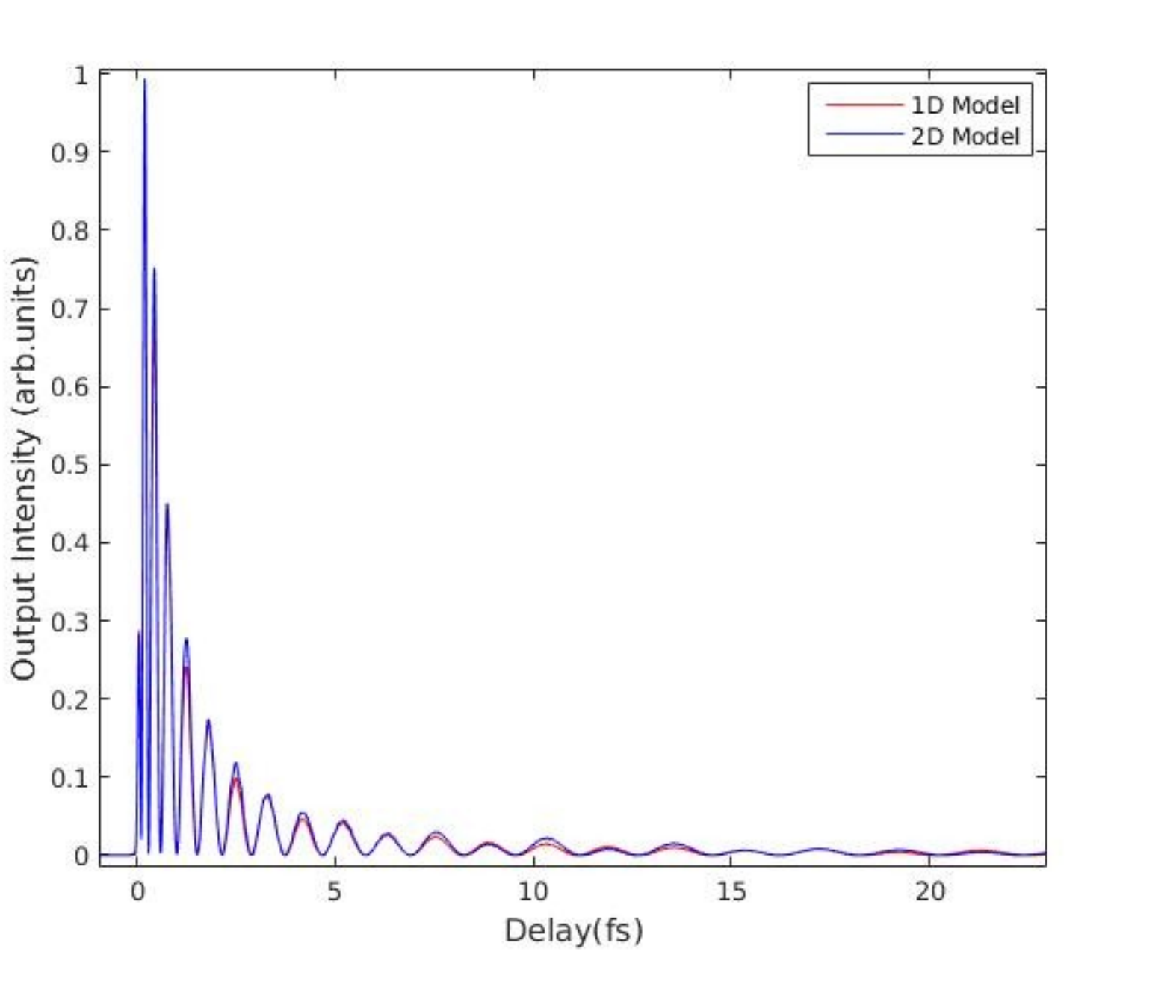}
	 \captionof{figure}{Propagated laser pulse intensity simulation in 1D model and 2D code.}
     \label{fig:Propagated laser pulse intensity simulation}
 \end{Figure}
\hspace{1mm}

\autoref{fig:Laser pulse intensity auto-correlation simulation} compares %
the laser pulse intensity autocorrelation of the models with that of the experimental measurement.  The error estimates represented by the error bars were computed by taking the standard deviation several autocorrelation samples divided by $\sqrt{N}$, where N was the number of samples we took at a given temperature. The computed autocorrelation of the models and that of the measurement are consistent with each other despite the models using a simplified spectrum and ignoring any nonlinear effects. Some of the nonlinear effects are apparent in \autoref{fig:Input and output spectrum}. Frequency dependent Kerr effect is most likely responsible for some of the spectral change, and a near-resonance frequency dependent Kerr model is currently under development. 

\begin{Figure}
     \centering
     \includegraphics[trim=2.5cm 15cm 3cm 2cm, clip=true,scale=0.45]{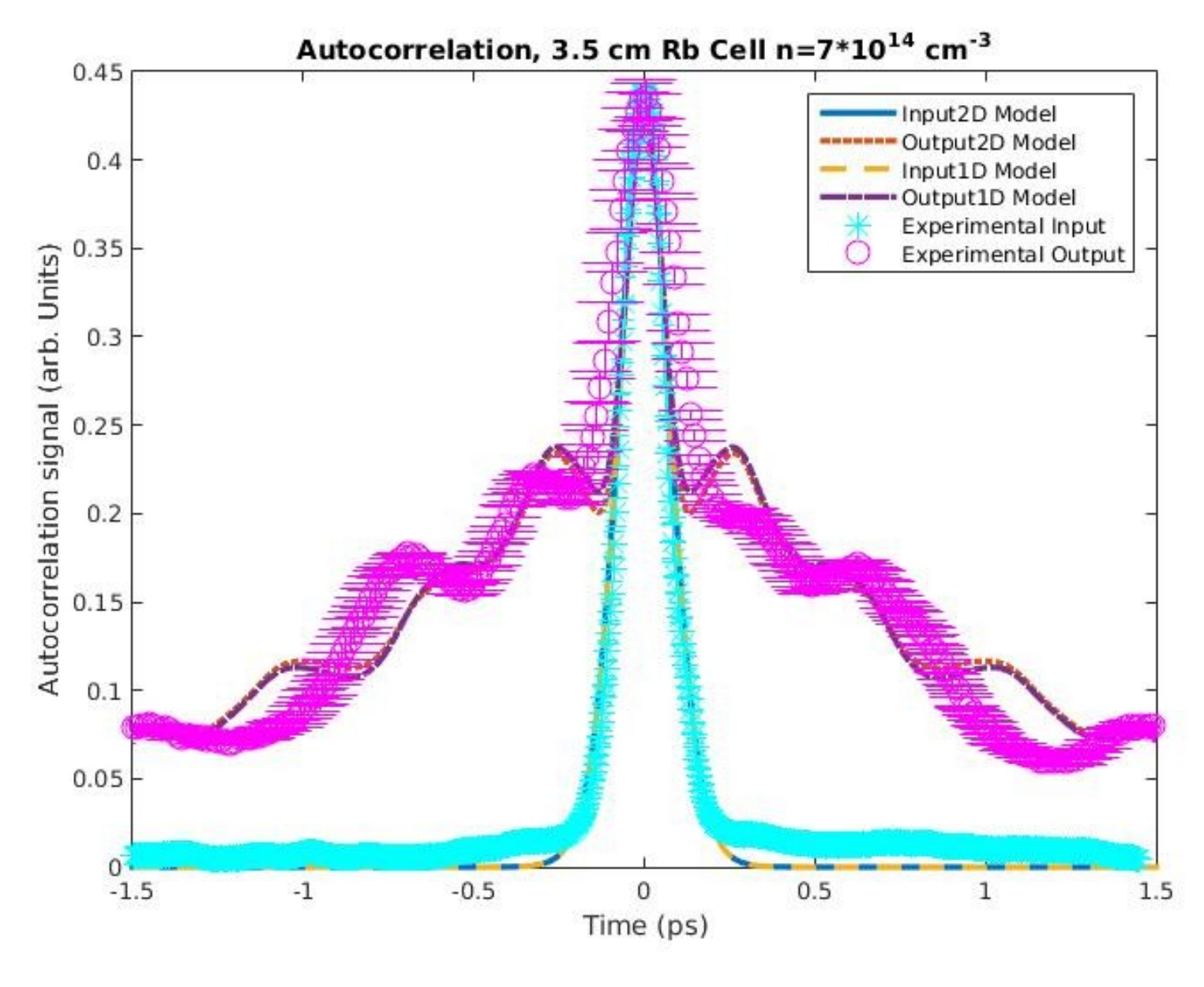}
	 \captionof{figure}{Laser pulse intensity autocorrelation simulation in 1D model, 2D code and experimental measurement.}
     \label{fig:Laser pulse intensity auto-correlation simulation}
 \end{Figure}
 \hspace{2mm}
 The 2D intensity simulation result and its auto correlation are consistent with the 1D model and experimental results that shows laser pulse stretching in the linear regime.

\autoref{fig:Diagram of Rb valence electronic state} shows that Rb has two absorption lines 
at 780.241 
and 794.979 nm 
from its ground state (see \autoref{fig:Input and output spectrum}).
Since these two lines are within the incoming laser pulse spectrum, strong excitation of electrons to the upper states of these transitions is expected.

\begin{Figure}
     \centering
     \includegraphics[trim=2cm 16cm 0cm 2cm, clip=true,scale=0.45]{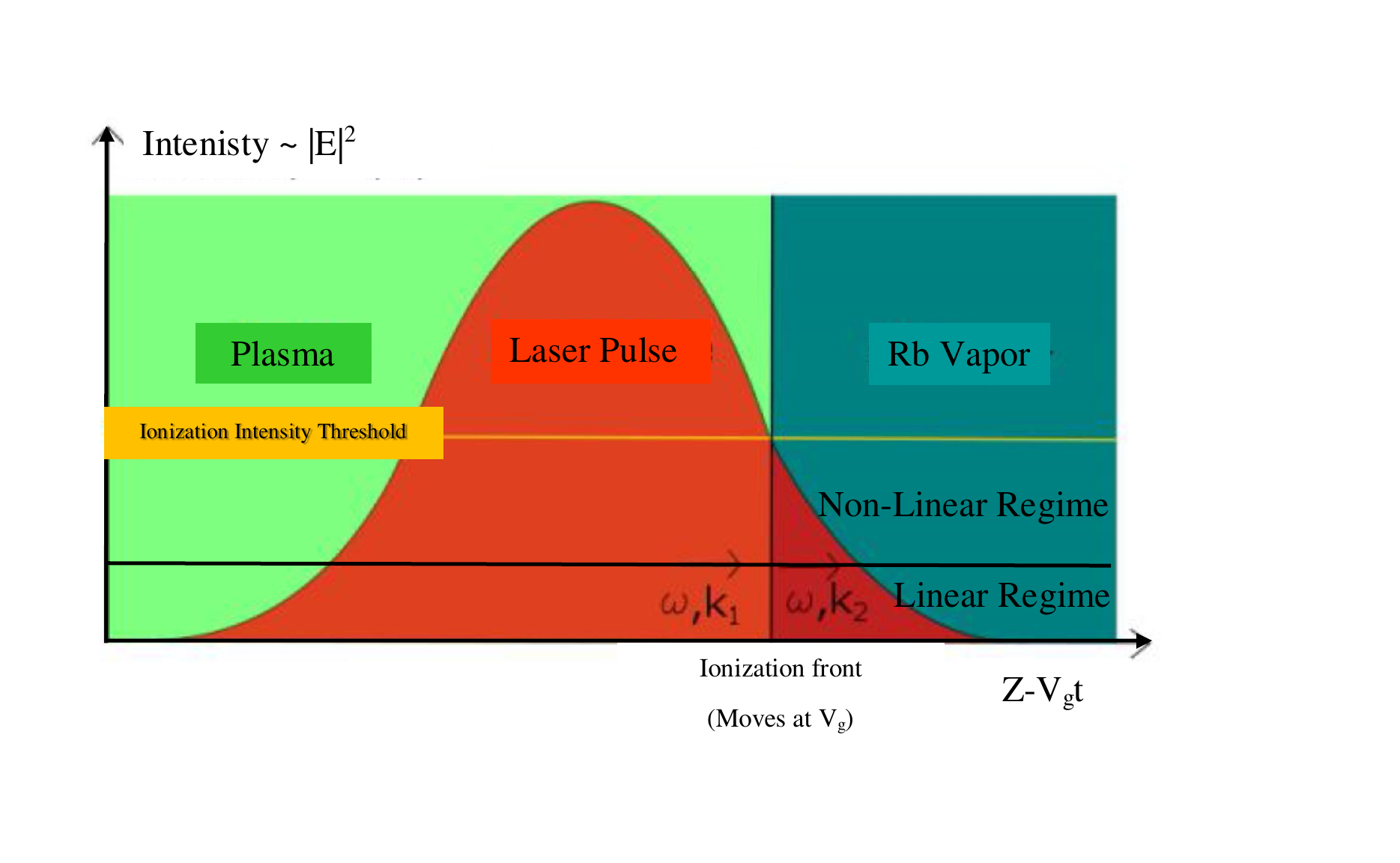}
	 \captionof{figure}{High intensity laser pulse.}
     \label{fig:High intensity laser pulse}
 \end{Figure} 
\hspace{4mm}

When the laser intensity is low, its interaction with the Rb vapor is the linear regime. 
 In this case the input and output 
 laser spectra are the same. 
 At higher laser pulse intensities 
 the interaction becomes nonlinear and the spectrum of laser pulse after the Rb vapor is expected to change. 
\begin{Figure}
     \centering
     \includegraphics[trim=2cm 14cm 0cm 2cm, clip=true,scale=0.45]{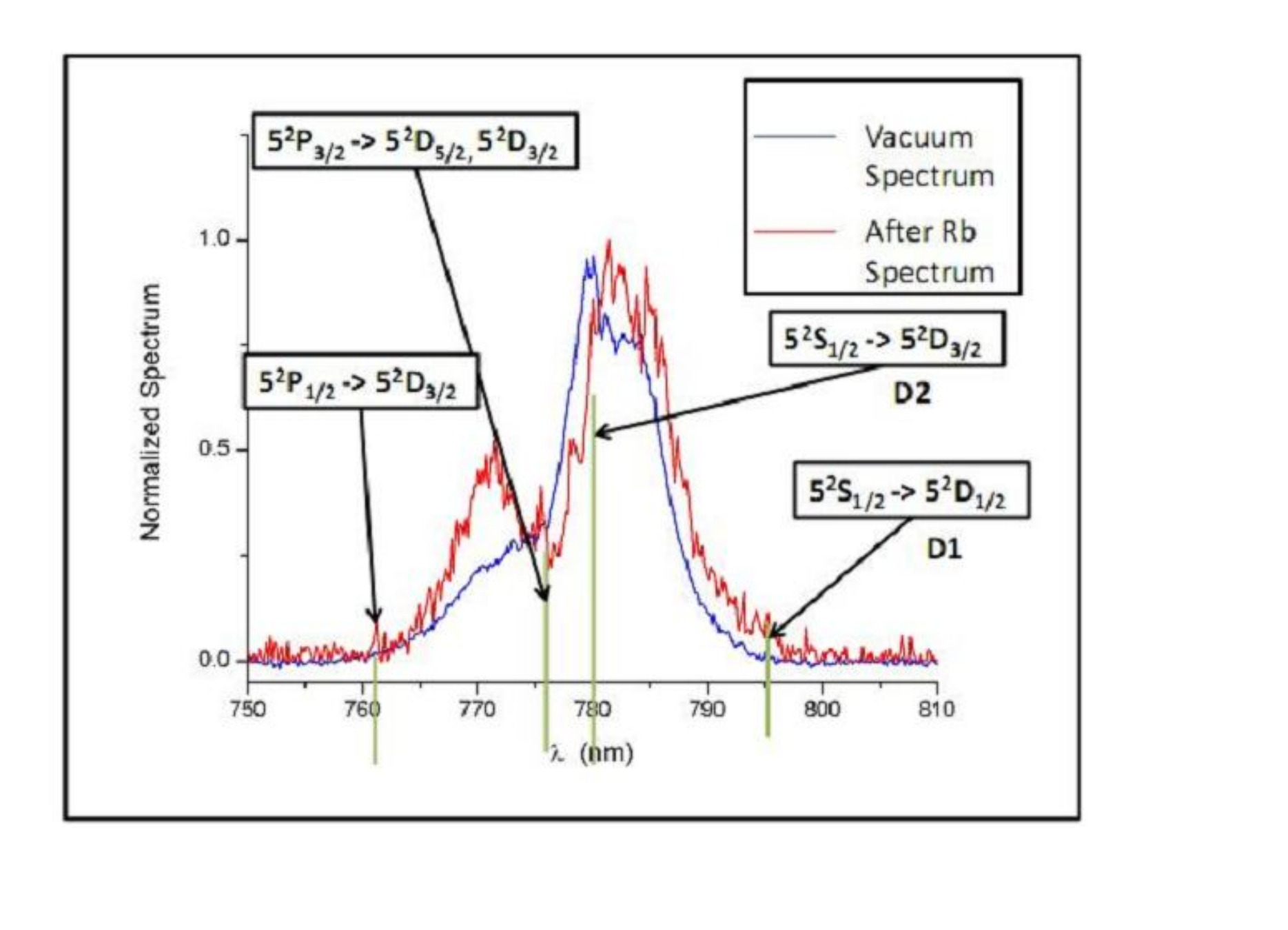}
	 \captionof{figure}{Expertimenral input and output spectrum through the 3.5~cm cell at Rb target with the Density $n\sim10^{15}~cm^{-3}$, I$\sim1~GW/cm^{2}$.}
    \label{fig:Input and output spectrum}
 \end{Figure} 
 \hspace{1mm}

For the AWAKE experiments the Rb vapor must be fully ionized (first electron) to transform the very uniform neutral Rb density into a very uniform plasma density. %
However, the laser pulse starts at low intensity and rises to above the ionization intensity with its $\sim$100 fs length (see \autoref{fig:High intensity laser pulse}). %
Therefore, various time slices of the laser pulse interact with the Rb vapor in various regimes. %
The very front, at low intensity (as described above) interacts in the linear regime. %
Other than narrow band absorption, its spectrum is thus expected not to change, but the head will stretch as we demonstrated in the experimental results presented here and to a much larger extend since the plasma length is 10 m. %
Later slices will interact in the nonlinear regime, they will also stretch in time and their spectrum will also broaden. %
Slices with intensities larger than the Rb ionization intensity will propagate in the plasma, with a much different dispersion relation ($\chi_{pe}=-\omega_{pe}^2/\omega_0^2$). %
In particular the plasma has a dispersion that is much smaller than that of the vapor (at the same density) and very weakly dependent on frequency. %
They will therefore propagate essentially without either stretching in time or spectral broadening. %
These slices will be able to ionize the Rb over long distances. %
We therefore expect that the anomalous dispersion within the ionizing laser pulse arising from the Rb D2 (and to some lesser extend D1) line will not prevent the propagation of the core of the laser at intensities large enough to ionize the vapor over the expected 10 m. %
In addition, the population of Rb upper atomic states by the strong pumping from the laser pulse broad spectrum should lead to more effective ionization. %
Indeed, field ionization relies on direct ionization from the ground state. %
The intensity required for this ionization process scales with the fourth power of the ionization potential. %
Therefore, any electron promoted to an upper Rb atomic state by the laser pulse through the D2, D1 and other lines (see \autoref{fig:Diagram of Rb valence electronic state}) will be ionized at much lower laser intensities. %
The interaction of the laser pulse with the vapor is of course more complicated than outlined here. %
In particular, time evolution along the laser pulse, as well and radial intensity variations must be included. %
We are therefore studying these effects with 2D numerical simulations and we perform corresponding experiments. %

\section{Conclusion}

We studied experimentally the effect of (linear) anomalous dispersion around the D2 and D1 lines of a Rb vapor on the propagation of a low-intensity short laser pulse. %
Experimental results are in good agreement with a simple 1D calculation and with the results of a 2D model that will be used to describe the propagation of a high intensity laser pulse that will be used to ionize the Rb vapor for the AWAKE experiment. %


\end{multicols}

\end{document}